\documentclass[dvipdfmx]{PoS}
\usepackage{amsmath,amssymb}
\usepackage{bm}
\usepackage{graphicx}

\title{Renormalization group flow of linear sigma model with $U_A(1)$
anomaly}

\ShortTitle{Renormalization group flow of linear sigma model with UA(1)
anomaly}

\author{\speaker{Tomomi Sato}\\
        High Energy Accelerator Research Organization (KEK), %
        Tsukuba 305-0801, Japan\\
        Graduate University for Advanced Studies (SOKENDAI), %
        Tsukuba 305-0801, Japan\\
        E-mail: \email{tomomis@post.kek.jp}}

\author{Norikazu Yamada\\
        High Energy Accelerator Research Organization (KEK), %
        Tsukuba 305-0801, Japan\\
        Graduate University for Advanced Studies (SOKENDAI), %
        Tsukuba 305-0801, Japan\\
        E-mail: \email{norikazu.yamada@kek.jp}}

\abstract{
 In the presence of finite $U_A(1)$ breaking, chiral phase transition of
 massless two-flavor QCD is studied by tracing the renormalization group
 flow of the corresponding effective theory.
 In the framework of the $\epsilon$ expansion, it is found that the nature of
 the phase transition depends on the initial condition for the parameters
 of the effective theory and that, when it undergoes second order phase
 transition, one of the universal exponents shows a different value from
 that in the standard $O(4)$ linear sigma model.
 We discuss that the origin of the difference is attributed to a
 non-decoupling effect.
 The present status of the calculation of the effective potential is
 presented.
}

\FullConference{The 32nd International Symposium on Lattice Field Theory,\\
		23-28 June, 2014\\
		Columbia University New York, NY}

\begin{document}

\section{Introduction}

Chiral phase transition at finite temperature $T_c$ is one of most
prominent features of QCD.
In 1980's, Psarski and Wilczek investigated the order of the phase
transition applying the $\epsilon$ expansion to linear sigma model
(LSM)~\cite{Pisarski:1983ms}, and pointed out that the nature of the
chiral phase transition of massless two-flavor QCD depends on the
existence of the flavor-singlet axial symmetry, $U_A(1)$, at the
critical temperature.

Two extreme cases have been well studied so far.
If the $U_A(1)$ breaking is infinitely large at $T_c$, the system would
be described by the standard $O(4)$ LSM and thus undergo second order
phase transition.
On the other hand, if the breaking vanishes at $T_c$~\cite{Aoki:2012yj},
the symmetry of the system is enhanced to $U(2)\times U(2)$.
In this case, the nature of the transition is still under debate.
It is argued that if it is of second the critical exponents in the
$U(2)\times U(2)$ universality class will
emerge~\cite{Pelissetto:2013hqa,Nakayama:2014}.

In this work, we investigate chiral phase transition of massless
two-flavor QCD with a finite $U_A(1)$ breaking.
Instead of dealing with two-flavor QCD directly, $U(2)\times U(2)$ LSM
with a $U_A(1)$ breaking, which we call the $U_A(1)$ broken LSM below,
is analyzed as the corresponding effective theory.
The nature is investigated through the renormalization group (RG) flow
calculated in the framework of the $\epsilon$ expansion.
Since the full results on this study are available in
Ref.~\cite{Sato:2014axa}\footnote{The early version is
Ref.~\cite{Sato:2013tka}}, we focus only on the highlights.
In the study of phase transitions, it is common to see effective
potential.
We present the preliminary result of the effective potential.


\section{RG flow of the $U_A(1)$ broken LSM}
\label{sec:review}

The LSM Lagrangian we study consists of the part preserving
$U(2)\times U(2)$ symmetry and the one breaking $U_A(1)$,
\begin{eqnarray}
      \mathcal{L}
&=&   \mathcal{L}_{U(2)\times U(2)}
    + \mathcal{L}_{\mathrm{breaking}},
 \label{eq:L_total}\\
      \mathcal{L}_{U(2)\times U(2)}
&=&   \frac{1}{2}\mathrm{tr}
      [\partial_{\mu} \Phi \partial_{\mu} \Phi^{\dagger}]
    + \frac{1}{2}{m_0}^2\mathrm{tr}[\Phi\Phi^{\dagger}]
    + \frac{\pi^2}{3}g_1\left(\mathrm{tr}[\Phi\Phi^{\dagger}]\right)^2
    + \frac{\pi^2}{3}g_2\mathrm{tr}\left[(\Phi\Phi^{\dagger})^2\right],
 \label{eq:L_0}\\
      \mathcal{L}_{\mathrm{breaking}}
&=& - \frac{c_A}{4}(\det\,\Phi+\det\,\Phi^{\dagger})
    + \frac{\pi^2}{3}x\,
      \mathrm{tr}[\Phi\Phi^{\dagger}](\det\,\Phi+\det\,\Phi^{\dagger})
    + \frac{\pi^2}{3}y\,(\det\,\Phi+\det\,\Phi^{\dagger})^2
\nonumber\\
& & + w \left(\mathrm{tr}
      \left[   \partial_\mu\Phi^{\dagger}t_2\partial_\mu\Phi^*t_2\right]
             + h.c.\right).
\label{eq:L_A}
\end{eqnarray}
where the building block
$\Phi= \sqrt{2}(\phi_0-i\chi_0)t_0 + \sqrt{2}(\chi_i+i\phi_i)t_i$ with
$t_0\equiv 1/2$ and $t_i\equiv \sigma_i/2$ transforms as
\begin{align}
 \Phi
 \to
  e^{2i\theta_A}L^{\dagger}\,\Phi\, R
  \
  (L\in SU_L(2),\, R\in SU_R(2),\, \theta_A\in \mathrm{Re}),
\end{align}
under chiral and $U_A(1)$ transformation.
Under the working hypothesis that this system undergoes second order
phase transition, the existence of an infrared-stable fixed point (IRFP) is
examined in $d=4-\epsilon$ dimension with $\epsilon=1$.

Rewriting eq.~(\ref{eq:L_total}) in terms of components, $\phi_a$ and
$\chi_a$, we obtain
\begin{align}
 \mathcal{L}_{\mathrm{total}}
 =&
  (1+w)\frac{1}{2}(\partial_\mu\phi_a)^2
  +\frac{1}{2}\left(m^2-\frac{c_A}{2}\right){\phi_a}^2
  +(1-w)\frac{1}{2}(\partial_\mu\chi_a)^2
  +\frac{1}{2}\left(m^2+\frac{c_A}{2}\right){\chi_a}^2
\notag \\
  &+\frac{\pi^2}{3}
     \left[
      \lambda({\phi_a}^2)^2
      +(\lambda-2x)({\chi_a}^2)^2
      +2(\lambda+g_2-z){\phi_a}^2{\chi_b}^2
      -2g_2(\phi_a\chi_a)^2
 \right],
 \label{eq:components}
\end{align}
where $\lambda=g_1+g_2/2+x+y,\ z=x+2y$ and $a$ runs 0 to 3.
In the following, $w=0$ is taken for simplicity.
Non-zero $c_A$ breaks degeneracy between $\phi_a$ and $\chi_a$ as
\begin{align}
 {m_{\phi}}^2=m^2-\frac{c_A}{2},
 \ \
 {m_{\chi}}^2=m^2+\frac{c_A}{2}.
\end{align}
We take $c_A>0$ and $m_{\phi}=0$ to realize QCD at $T_c$.
Then, $\chi_a$ has a mass of ${m_{\chi}}^2=c_A$.

It is naively expected that the massive fields $\chi_a$ decouple from
the system in the IR limit.
If it is the case, the system will eventually end up with well known
$O(4)$ LSM and hence undergo the second order phase transition
inhering in the $O(4)$ universality class.

In order to trace the effect of the massive fields, we take a mass dependent 
renormalization scheme\footnote{For two-point functions, the on-shell
scheme is applied, and hence $c_A$ does not run.}.
$\beta$ functions thus obtained are
\begin{align}
\label{eq:bl}
 \beta_{\hat\lambda}
 =
  \mu\frac{d\hat{\lambda}}{d\mu}
 =&
  -\epsilon \hat{\lambda}+2\hat{\lambda}^2
  +\frac{1}{6}f(\hat{\mu})
       (
        4\hat{\lambda}^2+6\hat{\lambda} \hat{g}_2+3{\hat{g}_2}^2
        -8\hat{\lambda}\hat{z}-6\hat{g}_2\hat{z}+4\hat{z}^2
       ),
\\
\label{eq:bg2}
 \beta_{\hat{g}_2}
 =
  \mu\frac{d\hat{g}_2}{d\mu}
 =&
  -\epsilon \hat{g}_2+\frac{1}{3}\hat{\lambda}\hat{g}_2
  +\frac{1}{3}f(\hat{\mu})\hat{g}_2(\hat{\lambda}-2\hat{x})
  +\frac{1}{3}h(\hat\mu)\hat{g}_2(4\hat{\lambda}+\hat{g}_2-4\hat{z}),
\\
\label{eq:bx}
 \beta_{\hat{x}}
 =
  \mu\frac{d\hat{x}}{d\mu}
 =&
  -\epsilon \hat{x}+4f(\hat\mu)(\hat{\lambda}\hat{x}-\hat{x}^2)
\notag \\ &
  +\frac{1}{12}(1-f(\hat\mu))
    (
     8\hat{\lambda}^2-6\hat{\lambda}\hat{g}_2
     -3{\hat{g}_2}^2+8\hat{\lambda}\hat{z}
     +6\hat{g}_2\hat{z}-4\hat{z}^2
    ),
\\
\label{eq:bz}
 \beta_{\hat{z}}
 =
  \mu\frac{d\hat{z}}{d\mu}
 =&
  -\epsilon \hat{z}
  +\frac{1}{2}(2\hat{\lambda}^2-\hat{\lambda}\hat{g}_2+2\hat{\lambda}\hat{z})
  -\frac{1}{6}h(\hat\mu)
    (4\hat{\lambda}^2-3{\hat{g}_2}^2-8\hat{\lambda}\hat{z}+4\hat{z}^2)
\notag \\&
  +\frac{1}{6}f(\hat\mu)
    (
     -2\hat{\lambda}^2+3\hat{\lambda}\hat{g}_2+3{\hat{g}_2}^2
     -2\hat{\lambda}\hat{z}-6\hat{g}_2\hat{z}+12\hat{\lambda}\hat{x}
     +6\hat{g}_2\hat{x}-12\hat{x}\hat{z}+4\hat{z}^2
    )
\end{align}
where the couplings with $\hat{}$ denote dimensionless couplings,
$\mu= \hat\mu\sqrt{c_A}$ is the renormalization scale, and
\begin{align}
 f(\hat{\mu})
 =
  1-\frac{4}{\hat{\mu}\sqrt{4+\hat{\mu}^2}} \arctan\frac{\hat{\mu}}{\sqrt{4+\hat{\mu}^2}},
 \ \ \
 h(\hat{\mu})
 =
  1-\frac{1}{\hat{\mu}^2} \log[1+\hat{\mu}^2],
\end{align}
with $f(\hat{\mu}\to 0)=\hat{\mu}^2/3,\,h(\hat{\mu}\to 0)=\hat{\mu}^2/2$, 
and $f(\hat{\mu}\to\infty)=h(\hat{\mu}\to\infty)=1$. 

The resulting RG flows can be classified into two cases, depending on
the asymptotic behavior of $\hat\lambda$ in the $\mu\to 0$ limit.
One case tends to be realized when $c_A$ is small, where all the
couplings diverge.
Thus, in this case the transition will be of first order.
The other case tends to happen when $c_A$ is relatively large.
In this case, all the couplings except for $\hat\lambda$ diverge, but
$\hat\lambda$ approaches $\hat\lambda_{\rm IRFP}=\epsilon/2$ that
coincides the IRFP of the $O(4)$ LSM.

The asymptotic behaviors of the diverging couplings are calculable and
turns out to be
\begin{align}
\label{eq:g2IR}
 \hat{g}_{2,\mathrm{asym}}(\mu)
 \equiv
  \lim_{\mu\to 0}\hat{g}_2(\mu)
 = c \hat\mu^{-\frac{5}{6}\epsilon},
\end{align}
\begin{align}
\label{eq:xIR}
 \hat{x}_{\mathrm{asym}}(\mu)
 \equiv
  \lim_{\mu\to 0}\hat{x}(\mu)
 =
  \frac{3}{32}\hat{g}_{2,\mathrm{asym}}^2(\mu),
\end{align}
\begin{align}
\label{eq:zIR}
 \hat{z}_{\mathrm{asym}}(\mu)
 \equiv
 \lim_{\mu\to 0}\hat{z}(\mu)
 =
  \frac{3}{4}\hat{g}_{2,\mathrm{asym}}(\mu),
\end{align}
where $c$ is a dimensionless constant depending on the initial conditions
of the RG flow.

With above results, we can calculate the approaching rate of $\hat\lambda$
to its IRFP.
Towards the infrared limit ($\hat\mu\to 0$),
$\beta_{\hat{\lambda}}$ becomes
\begin{align}
\label{eq:beta_lIR}
 \lim_{\mu\to 0}\beta_{\hat{\lambda}}
 =&
  -\epsilon \hat{\lambda}+2\hat{\lambda}^2
  +f(\hat\mu)
    \left(
     \frac{1}{2}\hat{g}_{2,\mathrm{asym}}^2
     -\hat{g}_{2,\mathrm{asym}}\hat{z}_{\mathrm{asym}}+\frac{2}{3}\hat{z}_{\mathrm{asym}}^2
    \right)+...
\notag \\
 =&
  -\epsilon \hat{\lambda}+2\hat{\lambda}^2
  +\frac{c^2}{24}\, \hat\mu^{1/3}.
\end{align}
Then, the approaching rate is found to be
\begin{eqnarray}
     \lim_{\mu\to 0}\hat\lambda(\mu) - \hat\lambda_{\rm IRFP}
\sim \hat\mu^{1/3}.
\end{eqnarray}

Usually, the approaching rate is discussed through the derivative of
the $\beta$ function,
 \begin{eqnarray}
  \omega
= \frac{d\beta_{\hat\lambda}}{d\hat\lambda}
  |_{\hat\lambda=\hat\lambda_{\rm IRFP}},
 \end{eqnarray}
which is one of the universal exponents\footnote{For exceptions, see
Ref.~\cite{Pelissetto:2000ek}.}, and the $O(4)$ LSM and the $U_A(1)$
broken LSM take
\begin{eqnarray}
 \omega_{O(4)}=\epsilon\ \ \ \mbox{and}\ \ \
 \omega_{U_A(1) {\rm broken}}=1-2\epsilon/3,
\end{eqnarray}
respectively.

General argument of renormalization group tells that $\omega$ is
determined by the RG dimension of the leading irrelevant operator.
In $O(4)$ LSM, $({\phi_a}^2)^2$ is the one.
In the $U_A(1)$ broken LSM, the leading irrelevant operator is not
evident but should not be the same as the $O(4)$ case because
$\omega_{O(4)}\ne \omega_{U_A(1) {\rm broken}}$.
To identify the leading irrelevant operator, we calculate $\omega$ with
$\hat{g}_2=0$ as a trial, which corresponds to omitting
$(\phi_a\chi_a)^2$ [see eq.(\ref{eq:components})].
Then, $\omega=\epsilon$ is obtained.
From this result, it is concluded that the operator $(\phi_a\chi_a)^2$
is the leading irrelevant operator in the $U_A(1)$ broken LSM.
Notice that this operator is not invariant under the $O(4)$ rotation in
the $O(4)$ LSM, which acts only on $\phi_a$ but leaves $\chi_a$
unchanged.

The above discussion may sound strange because, if the decoupling
theorem~\cite{{Symanzik:1973vg},Appelquist:1974tg} holds in this system,
the massive field $\chi_a$ should not affect the low energy behavior of
the system.
However, the decoupling theorem becomes non-trivial when a theory
contains a dimensionful coupling~\cite{Aoki:1997er}.

Since one of the universal exponents, $\omega$, turns out to be
different from that in the standard $O(4)$ LSM, it is clearly
interesting to see the other exponents ($\nu$ and $\eta$ in the standard
notation).
We have calculated these exponents in the $U_A(1)$ broken LSM through
one-loop, and found that they are the same as those in the $O(4)$ LSM.
Two-loop calculation is ongoing.

Whether $\hat\lambda$ flows into $\hat\lambda_{\rm IRFP}$ or not depends
on not only $c_A$ but also the initial conditions for the couplings.
We have determined for various values of $c_A$ the parameter space of
the initial condition (attractive basin) with which
$\hat\lambda\to\hat\lambda_{\rm IRFP}$.
It is found that the attractive basin shrinks as $c_A$ decreases.

The scheme dependence of the low energy behavior was also examined by
taking the $\overline{\rm MS}$ scheme as an alternative.
Since the $\overline{\rm MS}$ scheme does not take care of the finite
mass of $\chi_a$ and hence the $\beta$ function does not feel the
decoupling, we compared the low energy behavior of the four-point
Green's function of $\phi_a$ between different schemes.
No scheme dependence is observed at the one-loop level.
For details of the above considerations, see Ref.~\cite{Sato:2014axa}.

\section{Effective potential}
\label{sec:Effpot}

In order to gain more understanding, we are calculating the effective
potential of the $U_A(1)$ broken LSM.
Our setup indicated by the working hypothesis and $m_{\phi}=0$ corresponds
to the system slightly above $T_c$.
Thus the order parameter $\langle\Phi\rangle$ must be zero.
By looking at the effective potential, we check whether this is indeed
the case.
In the following, we concentrate on the case where
$\langle \chi_a \rangle= 0$ for simplicity.
The calculation of the general case is not completed yet.
We choose the vacuum expectation value as
$\langle\phi_a\rangle=\{\phi_{cl},0,0,0\}$.
The explicit one-loop calculation yields the effective potential,
\begin{align}
 V_{\mathrm{eff}}(\phi_{cl})/\mu^{\epsilon}
 =
  \frac{\pi^2}{3}\hat{\lambda}(\mu){\phi_{cl}}^4
  +V_{\phi}^{\mathrm{1-loop}}(\phi_{cl})/\mu^{\epsilon}
  +V_{\chi}^{\mathrm{1-loop}}(\phi_{cl})/\mu^{\epsilon}.
\end{align}
where the first term represents the tree level result, and the second
and third terms are the one-loop contributions from the diagrams
containing $\phi_a$ and $\chi_a$ in the loop, respectively. 
Neglecting $V^{\rm 1-loop}_{\chi}(\phi_{cl})$, $V_{eff}$ agrees with
that of $O(4)$ LSM.
The explicit expressions for the one-loop contributions are
\begin{eqnarray}
    V_{\phi}^{\mathrm{1-loop}}(\phi_{cl})/\mu^{\epsilon}
&=& \frac{\pi^2}{3} \hat{\lambda}^2(\mu)
    \left(
          \log\left[\frac{4}{3}\pi^2\hat{\lambda}(\mu)\frac{{\phi_{cl}}^2}{\mu^{2-\epsilon}}\right]
        + \frac{1}{2}+\frac{3}{4}\log 3
    \right) {\phi_{cl}}^4,
\\
\label{eq:Vchi}
    V_{\chi}^{\mathrm{1-loop}}(\phi_{cl})/\mu^{\epsilon}
&=& \frac{\mu^{-2\epsilon}}{4(4\pi)^2}
    \Biggl[ \,
       (m_1^2)^2 \log[m_1^2/c_A]
     + 3\,(m_2^2)^2 \log[m_2^2/c_A]
\nonumber\\
 &&\hspace{8ex}
  - \left(\frac{4}{3}\pi^2\,\mu^{\epsilon}\right)^2
  \{(\hat{\lambda}(\mu)-\hat{z}(\mu))^2
  +3(\hat{\lambda}(\mu)+\hat{g}_2(\mu)-\hat{z}(\mu))^2\}
\nonumber\\
& & \hspace{10ex}\times
    \left(   \int^1_0d\xi \log\left[1+\xi(1-\xi)\frac{\mu^2}{c_A}\right]
           + \frac{3}{2}
    \right) {\phi_{cl}}^4
\nonumber\\
 & & \hspace{8ex}
  - \frac{4}{3}\pi^2\, \mu^{\epsilon}
       \{(\hat{\lambda}(\mu)-\hat{z}(\mu))
     + 3(\hat{\lambda}(\mu)+\hat{g}_2(\mu)-\hat{z}(\mu))\}
       c_A{\phi_{cl}}^2
   \Biggr], 
\end{eqnarray}
where
\begin{eqnarray}
 m_1^2
 \equiv
  \frac{4}{3}\pi^2\,\mu^{\epsilon}(\hat{\lambda}(\mu)-\hat{z}(\mu)){\phi_{cl}}^2+c_A,
\ \ \
 m_2^2
 \equiv
  \frac{4}{3}\pi^2\,\mu^{\epsilon}
   (\hat{\lambda}(\mu)+\hat{g}_2(\mu)-\hat{z}(\mu)){\phi_{cl}}^2+c_A.
\end{eqnarray}
By solving the RG equation
\begin{align}
 \left[
  \mu\frac{\partial}{\partial\mu}
  +\sum_i\beta_{\hat{g}_i}\frac{\partial}{\partial\hat{g}_i}
  -\gamma\phi_{cl}\frac{\partial}{\partial\phi_{cl}}
 \right]
  V_{\mathrm{eff}}(\phi_{cl})
 =
  0,
\end{align}
we implement the RG improvement, which is realized by simply setting
$\mu={\phi_{cl}}^{2/(2-\epsilon)}$ in the unimproved expressions.
Where $\{\hat{g}_i\}=\{\hat\lambda,\hat{g}_2,\hat{x},\hat{z}\}$. 
Because there are two scales $\phi_{cl}$ and $c_A$, 
we need more powerful technique for the resummation. 
Owing to the RG improvement, we can easily see that the effect of 
$V_{\phi}^{\mathrm{1-loop}}$ is just a shift of the coefficient of the tree potential. 
In the attractive basin, $\hat{\lambda}(\mu)$ goes to the positive constant $\epsilon/2$
in $\mu\to 0$.
Therefore, $V_{\phi}^{\mathrm{1-loop}}$ does not produce a non-trivial minimum 
as seen in the numerical calculation. 

\begin{figure}[tb]
 \begin{center}
  \includegraphics[width=0.7 \textwidth]{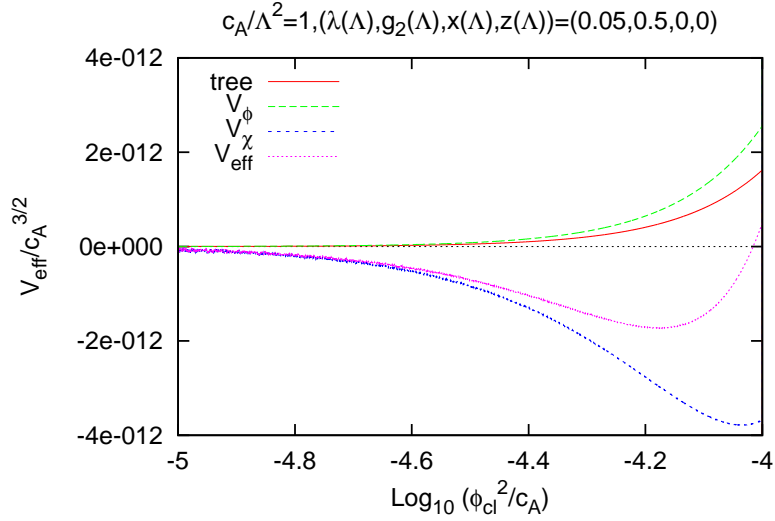}
 \end{center}
 \caption{Effective potential and contributions from each term. %
 Parameters are set to be $c_A/\Lambda^2=1$ and %
 $(\hat\lambda(\Lambda),\hat{g}_2(\Lambda),\hat{x}(\Lambda),\hat{z}(\Lambda))=%
 (0.05,0.5,0,0)$. %
 This is inside of the attractive basin. $\epsilon$ is set to 1. %
 There is a non-trivial minimum produced by higher order contribution in %
 $V_{\chi}^{\mathrm{1-loop}}$.}
 \label{fig:Veff}
\end{figure}

Fig.~\ref{fig:Veff} shows shapes of the RG improved effective potential 
drawn by numerical calculation. 
The largest contribution near the origin comes from $V_{\chi}^{\mathrm{1-loop}}$, 
and it produces the non-trivial minimum. 
In this region, the loop contribution is larger than the tree contribution, 
the perturbative calculation is disputable. 
To get more information about the minimum, we estimate the behavior of 
$V_{\chi}^{\mathrm{1-loop}}$ near the origin. 
Since $\hat{g}_2{\phi_{cl}}^2$ and $\hat{z}{\phi_{cl}}^2$ converge to zero 
in $\mu={\phi_{cl}}^{2/(2-\epsilon)}\to 0$, we can expand
$V_{\chi}^{\mathrm{1-loop}}$ around $\phi_{cl}=0$ as
\begin{align}
\label{eq:Vchi_ap}
 V_{\chi}^{\mathrm{1-loop}}(\phi_{cl})/\mu^{\epsilon}
 \approx&
  \frac{\pi^4}{9} \mu^{\epsilon} 
   \{(\hat{\lambda}-\hat{z})^3+3(\hat{\lambda}+\hat{g}_2-\hat{z})^3\}
   \frac{{\phi_{cl}}^6}{c_A}
\notag \\&
  -\frac{\pi^2}{6^3}\{(\hat{\lambda}-\hat{z})^2+3(\hat{\lambda}+\hat{g}_2-\hat{z})^2\}
      {\phi_{cl}}^4 \frac{\mu^2}{c_A}
   +...\,,
\end{align}
where the first term comes from $(m_1^2)^2\log[m_1^2/c_A]$ and $(m_2^2)^2\log[m_2^2/c_A]$
in eq.(\ref{eq:Vchi}). 
These terms have linear and quadratic in couplings , 
though they are canceled by other terms. 
The second term comes from the one including $\log[1+\xi(1-\xi)\mu^2/c_A]$. 
Using the IR asymptotic behaviors of couplings (eq.(\ref{eq:g2IR})-(\ref{eq:zIR})), 
\begin{align*}
 \lim_{\mu\to 0} \mu^{\epsilon}
  \{(\hat{\lambda}-\hat{z})^3+3(\hat{\lambda}+\hat{g}_2-\hat{z})^3\}
  \frac{{\phi_{cl}}^6}{c_A}
 =&
  -\frac{3}{8}c^3c_A^{\frac{5}{4}\epsilon-1}\mu^{-\frac{3}{2}\epsilon}{\phi_{cl}}^6
 =
  -\frac{3}{8}c^3c_A^{\frac{5}{4}\epsilon-1}{\phi_{cl}}^{6-3\epsilon},
\\
 \lim_{\mu\to 0}
  \{(\hat{\lambda}-\hat{z})^2+3(\hat{\lambda}+\hat{g}_2-\hat{z})^2\}
   {\phi_{cl}}^4\frac{\mu^2}{c_A}
 =&
  \frac{5}{8}c^2c_A^{\frac{5}{6}\epsilon-1}\mu^{2-\frac{5}{3}\epsilon}{\phi_{cl}}^4
 =
  \frac{5}{8}c^2c_A^{\frac{5}{6}\epsilon-1}{\phi_{cl}}^{8-\frac{10}{3}\epsilon}.
\end{align*}
In the rightmost in each line, we use $\mu={\phi_{cl}}^{2/(2-\epsilon)}$. 
The first term of eq(\ref{eq:Vchi_ap}) is proportional to ${\phi_{cl}}^3$ 
and the second term is proportional to ${\phi_{cl}}^{4+\frac{2}{3}}$ in $\epsilon\to 1$, 
The first term gives larger contribution than ${\phi_{cl}}^4$. 
This term becomes negative when $c>0$, therefore another minimum other
than $\phi_{cl}=0$ exists as shown in the numerical calculation. 
However, this contribution is cubic in couplings.
Because we carry out the calculations to quadratic order, the cubic
order is subject to unknown two-loop coefficients.
To know whether the non-trivial minimum arises or not, the two-loops
calculation is needed.

\section{Summary}

In the $\epsilon$ expansion, we calculated the RG flow and the effective
potential of the $U_A(1)$ broken LSM.
There are two cases of the flow depending on size of the $U_A(1)$ breaking. 
With small $U_A(1)$ breaking, the flow tends to diverge in the IR limit. 
Increasing the breaking, one of the coupling converges to the same value with 
the IR fixed point of the $O(4)$ LSM. 
In this case, the critical exponents, $\omega$, $\nu$, $\eta$ are
calculated to one-loop.
$\omega$ turns out to differ from that in $O(4)$, while the others are
found to be the same to one-loop order. 
This difference comes from the leading irrelevant operator in the
$U_A(1)$ broken LSM and can be interpreted as the non-decoupling effect.
We determined the attractive basin in parameter space of the $U_A(1)$
broken LSM.
This attractive basin shrinks as the size of the $U_A(1)$ breaking decreasing. 
There is a non-trivial minimum in the RG improved effective potential. 
However, this minimum is produced by the order in couplings higher than we have performed. 
We need to carry out the higher loops calculation to establish the location of 
the minimum.

\end{document}